\documentclass[aps,prb, reprint, superscriptaddress, longbibliography]{revtex4-1}

\usepackage{graphicx}
\usepackage[hidelinks]{hyperref}

\usepackage{color}
\usepackage{soul}
\usepackage{dcolumn}
\usepackage{amsmath}
\usepackage{amssymb}
\usepackage{bm}

\begin{document}

\title{Fluorescence shelving of a superconducting circuit}
\title{Cavityless circuit quantum electrodynamics }
\title{Electron shelving of a superconducting artificial atom}

\author{Nathana\"el Cottet}%
\thanks{These two authors contributed equally}
\affiliation{Physics Department, Joint Quantum Institute, and Quantum Materials Center, University of Maryland, College Park, MD 20742}
\affiliation{%
	Universit\'e Lyon, ENS de Lyon, Universit\'e Claude Bernard Lyon 1, CNRS, Laboratoire de Physique, F-69342 Lyon, France
}

\author{Haonan Xiong}
\thanks{These two authors contributed equally}
\affiliation{Physics Department, Joint Quantum Institute, and Quantum Materials Center, University of Maryland, College Park, MD 20742}

\author{Long Nguyen}
\affiliation{Physics Department, Joint Quantum Institute, and Quantum Materials Center, University of Maryland, College Park, MD 20742}

\author{Yen-Hsiang Lin}
\affiliation{Physics Department, Joint Quantum Institute, and Quantum Materials Center, University of Maryland, College Park, MD 20742}

\author{Vladimir Manucharyan}
\affiliation{Physics Department, Joint Quantum Institute, and Quantum Materials Center, University of Maryland, College Park, MD 20742}
\date{\today}

\maketitle

\textbf{Interfacing stationary qubits with propagating photons is a fundamental problem in quantum technology~\cite{kimble2008quantum}. Cavity quantum electrodynamics~\cite{haroche2006exploring} (CQED) invokes a mediator degree of freedom in the form of a far-detuned cavity mode, the adaptation of which to superconducting circuits (cQED) proved remarkably fruitful~\cite{wallraff2004strong, haroche2020cavity, blais2020quantum}. The cavity both blocks the qubit emission and it enables a dispersive readout of the qubit state. Yet, a more direct (cavityless) interface is possible with atomic clocks, in which an orbital cycling transition can scatter photons depending on the state of a hyperfine or quadrupole qubit transition~\cite{ludlow2015optical}. Originally termed ``electron shelving"~\cite{nagourney1986shelved}, such a conditional fluorescence phenomenon is the cornerstone of many quantum information platforms, including trapped ions~\cite{bruzewicz2019trapped}, solid state defects~\cite{robledo2011high, sukachev2017silicon}, and semiconductor quantum dots~\cite{vamivakas2009spin}. Here we apply the shelving idea to circuit atoms and demonstrate a 
conditional fluorescence	
readout of
fluxonium qubit~\cite{Manucharyan113} 
placed inside a matched one-dimensional waveguide~\cite{astafiev2010resonance}. Cycling the non-computational transition between ground and third excited states produces a microwave photon every $91~\textrm{ns}$ conditioned on the qubit ground state, while the qubit coherence time exceeds $50~\mu s$. The readout has a built-in quantum non-demolition property~\cite{braginsky1996quantum}, allowing over 100 fluorescence cycles in agreement with a four-level optical pumping model. 
Our result introduces a resource-efficient alternative to cQED. It also adds a state-of-the-art quantum memory to the growing toolbox of waveguide QED~\cite{hoi2011demonstration, van2013photon,mirhosseini2019,kannan2020}. 
}

An important first step towards eliminating the cavity was accomplished with the observation of resonant fluorescence by a flux qubit directly coupled to the one-dimensional vacuum of a matched $50~\Omega$ transmission line~\cite{astafiev2010resonance}. Resonance fluorescence was also explored with transmons in relation to several fundamental topics in quantum optics~\cite{houck2007generating, campagne2014observing, toyli2016resonance, cottet2017}. On one hand, circuit atoms placed inside microwave transmission lines routinely reach a nearly unit light-matter coupling efficiency, i.e. the rate of spontaneous emission into the waveguide largely surpasses the non-radiative decoherence. Achieving such regime is extremely difficult in conventional quantum optics. On the other hand, the lack of diversity in transition frequencies and selection rules prevented a full-blown quantum control of transmons and flux qubits outside the ``qubit-in-a-cavity" paradigm of cQED~\cite{englert2010mesoscopic}.

\begin{figure*}[ht]
	\centering
	\includegraphics[width=\linewidth]{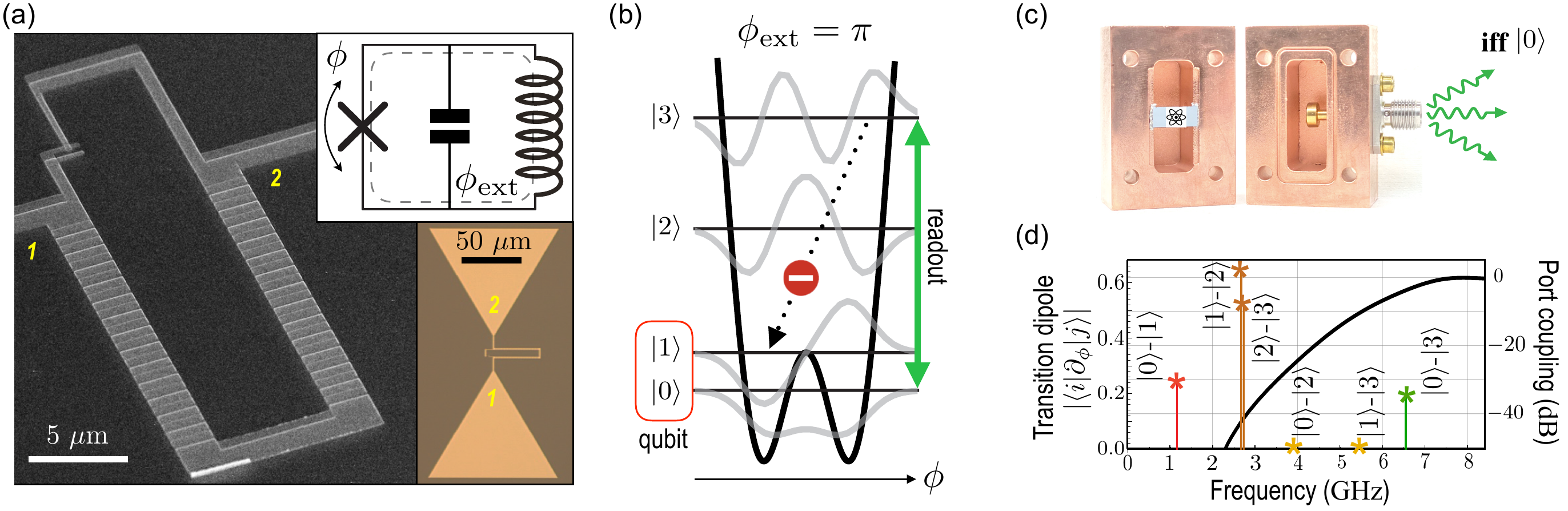}
	\caption{\textbf{Cavityless circuit quantum electrodynamics.} (a) Scanning electron microscope image of the fluxonium's superconducting loop, made of one weak Josephson junction and an array of stronger junctions. Top inset: fluxonium's effective three elements circuit model. Bottom inset: the weak junction is connected to a capacitive bow-tie antenna (optical image). (b) Effective potential seen by the phase-difference $\phi$ across the weak junction (black), energy levels (horizontal black lines) and wavefunctions (gray) at the half-integer external flux through the loop, $\phi_{\textrm{ext}} = \pi$. 
		The qubit levels are $|0\rangle$ and $|1\rangle$, the readout transition 
		is $|0\rangle - |3\rangle$, and even transitions are dipole-forbidden. (c) Coaxial-to-waveguide adapter for broadband coupling of the on-chip circuit to a $50~\Omega$-cable. (d) Measured port coupling (solid black) vs. frequency and the calculated transition dipoles
		 (colored stars). Note the absence of resonant modes in the adapter in the entire $0-8~\textrm{GHz}$ frequency range.
	}
	\label{fig1}
\end{figure*}

Our cavityless experiment is made possible by a unique combination of strong anharmonicity and long coherence time of modern fluxonium qubits~\cite{nguyen2019high, zhang2020universal}. Fluxonium consists of a weak Josephson junction shunted by an appropriately high-value inductance and a capacitance, here in the form of bow-tie antennas  (Fig.~\ref{fig1}a). When cooled-down near 10~mK in a dilution refrigerator the circuit dynamics can be characterized by a single collective degree of freedom, the superconducting phase-difference $\phi$ across the weak junction.  At the half-integer magnetic flux through the loop, $\phi_{\textrm{ext}} = \pi$, the variable $\phi$ behaves like the coordinate of a particle inside a symmetric double-well potential and the spectrum is insensitive to flux noise in the first order (Fig.~\ref{fig1}b). Tunneling of $\phi$ across the barrier forms a qubit transition between the two lowest energy eigenstates $|0\rangle$ and $|1\rangle$ at frequency $\omega_{01} = 2\pi \times 1.152~\textrm{GHz}$. The non-computational states $|2\rangle$, $|3\rangle$, etc., correspond to one-dimensional orbitals of $\phi$ at energies above the barrier. In particular, transition $|0\rangle$-$|3\rangle$ at frequency $\omega_{03}= 2\pi\times 6.544~\textrm{GHz}$ suits well for fluorescence cycling~\cite{supp}. Note that such transition is allowed, unlike in transmons, thanks to the absence of a harmonic ladder selection rule in our non-linear circuit.

Shelving requirements were met by combining the parity selection rules at $\phi_\mathrm{ext}=\pi$ with a properly engineered microwave environment. We mount the device chip inside a rectangular enclosure with a specially designed single input/output port (Fig.~\ref{fig1}c). The port can be viewed as a broadband coaxial-to-waveguide adapter~\cite{kou2018simultaneous}. Above the cut-off frequency, given by the enclosure dimensions, the port coupling is near unity in a few-$\textrm{GHz}$ band; below the cut-off, the port coupling drops rapidly with frequency (Fig.~\ref{fig1}d, solid line). Therefore, the radiative lifetime of state $|1\rangle$ can be orders of magnitude longer than that of state $|3\rangle$, even though transitions $|0\rangle$-$|1\rangle$ and $|0\rangle$-$|3\rangle$ have comparable dipoles -- matrix elements of the Cooper pair number operator $-i\partial_{\phi}$ (Fig.~\ref{fig1}d). Moreover, radiative decays $|3\rangle \rightarrow |1\rangle$ and $|3\rangle\rightarrow |2\rangle \rightarrow |1\rangle$ are inhibited by the parity selection rule and the weak port coupling, respectively. 
As a result, resonant driving of $|0\rangle$-$|3\rangle$ transition would continuously generate fluorescence if and only if the atom (and hence the qubit) is initially in state $|0\rangle$. Despite a periodic transit of population through the non-computational state $|3\rangle$, the atom rapidly relaxes back to state $|0\rangle$ on switching off the drive. Likewise, the atom starting in state $|1\rangle$ remains unperturbed by the far off-resonant readout drive and hence our conditional fluorescence readout has a built-in QND property. 

Following previous work~\cite{astafiev2010resonance, hoi2011demonstration}, we detect fluorescence using a phase-sensitive microwave reflectometry setup. For the time being, let us neglect the possibility of transitions outside the cycling manifold $\{|0\rangle, |3\rangle\}$, and assume the environment to be at zero temperature. According to the input/output theory, the complex amplitude of the outgoing reflected wave $\alpha_{out}$ is given by the sum of the incoming wave amplitude $\alpha_{in}$ and the fluorescence field emitted by the atom\cite{gardiner1985}. For a drive near the readout frequency $\omega_{03}$, this relation takes the simple expression $\alpha_{out} = \alpha_{in} -\sqrt{\Gamma} \langle |0\rangle \langle 3|\rangle$, where $\Gamma$ is the rate of direct radiative decay $|3\rangle\rightarrow|0\rangle$, and the expectation value is taken over the dissipative driven steady state. The fluorescence contribution is zero if the atom is not initially in the ground state, which allows linking the complex reflection amplitude $r = \alpha_{out}/\alpha_{in}$ to the ground state population $p_0$:
\begin{equation}
\label{equ1}
r=1-2p_0\times \frac{\Gamma^2/2-i(\omega-\omega_{03})\Gamma}{\Gamma^2/2+\Omega^2+2(\omega-\omega_{03})^2}
\end{equation}
with $\omega$ the drive frequency. Here the Rabi rate $\Omega = 2\sqrt{\Gamma}\alpha_{in}$ conveniently represents the readout drive strength in units of $[\textrm{Hz}]$. The information about the state of the atom is carried by the fluorescence signal, hence is proportional to $\Omega|1-r(\omega, \Omega)|$. It is maximal when the drive amplitude and frequency obey $\Omega = \Gamma/\sqrt{2}$ and $\omega = \omega_{03}$, which reduces Eq.~(\ref{equ1}) to $r = 1-p_0$. Note that phase-sensitive fluorescence detection relies on the drive-induced coherence between states $|0\rangle$ and $|3\rangle$, rather than on the population of state $|3\rangle$ in the more common case of optical photodetection. 

\begin{figure*}[ht]
	\centering
	\includegraphics[width=0.6\linewidth]{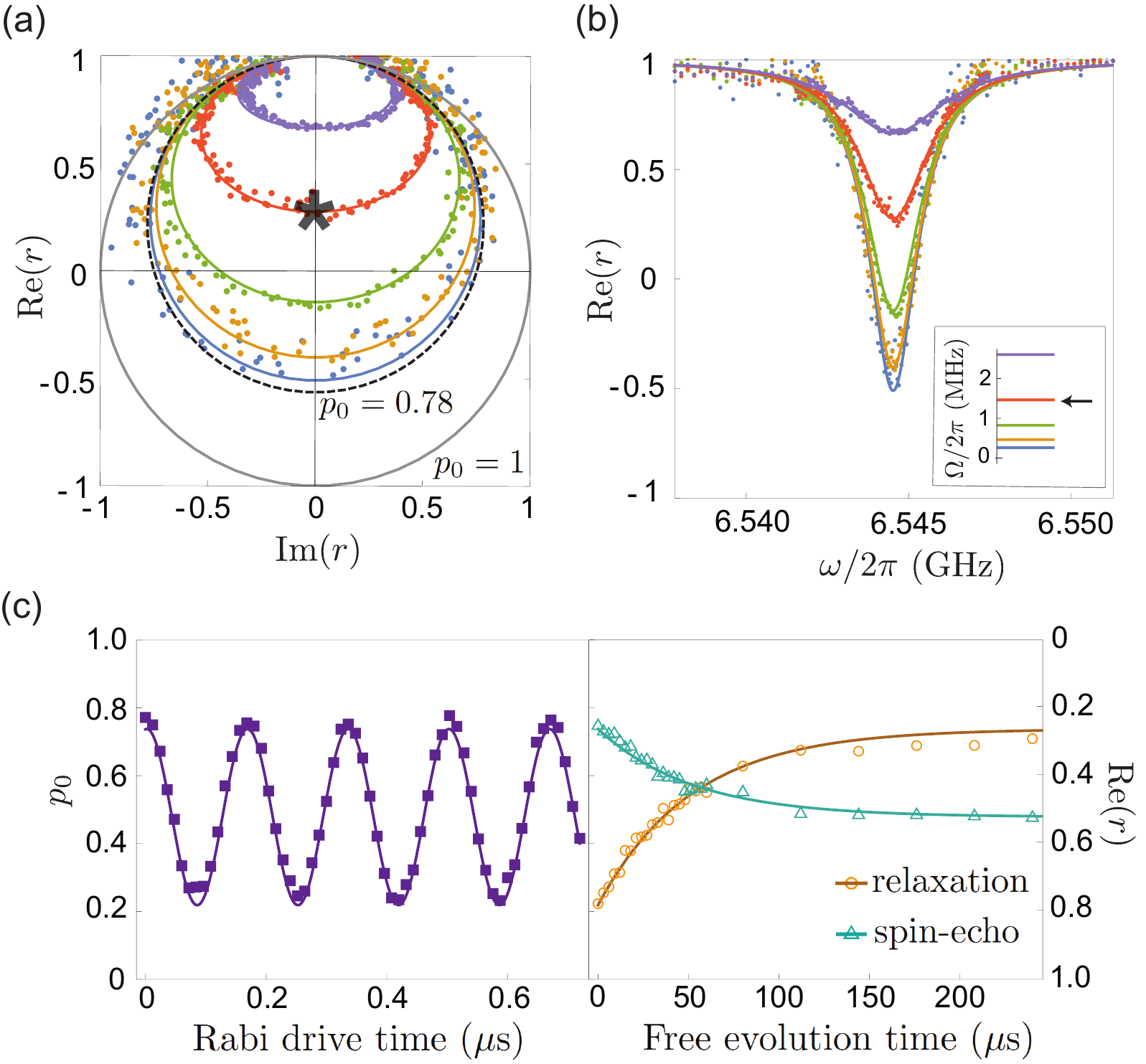}
	\caption{\textbf{Conditional fluorescence readout.} (a) Complex plane representation and (b) real part of the reflection coefficient $r$ as a function of readout frequency $\omega$ at various powers represented by the $|0\rangle$-$|3\rangle$ transition Rabi frequency $\Omega$ (inset). The experimental data (colored dots) are fitted by the theory of Eq.~(\ref{equ1}) (lines) with $\Gamma=2\pi\times 1.74$~MHz and $p_0=0.78$ due to thermal occupation of excited states. In the limit $\Omega\rightarrow0$, the data in Fig.~2a becomes a circle with a radius $p_0$ (gray circle, $p_0 =1$ and dashed circle, $p_0 = 0.78$). The optimal readout settings are highlighted by the black star and arrow (see text). (c) Rabi oscillations (left panel), energy relaxation (right panel, yellow circles), and spin-echo coherence decay (green triangles) measured using conditional fluorescence readout. Fitting to exponential decay yields (coincidentally) $T_1=T_2 =52\ \mu\textrm{s}$.
	}
	\label{fig2}
\end{figure*}

We begin our experiment by spectroscopically probing the readout transition (Fig.~\ref{fig2}a,b). The measured reflection amplitude $r$ as a function of drive frequency $\omega$ and amplitude $\Omega$ fits well to the Eq.~(\ref{equ1}) model. Only four adjustable parameters are used in the global fit: $\omega_{03} =2\pi\times 6.544~\textrm{GHz}$, $\Gamma = 2\pi \times 1.74~\textrm{MHz}$, the thermal equilibrium ground state population $p_0^{\textrm{th}} = 0.78$, and the scaling factor converting the value of $\Omega$ to the RF-source amplitude. We integrate the reflection signal during a time $\tau_m = 0.5~\mu\textrm{s}$ after a delay $\tau_w = 0.5~\mu\textrm{s}$. The wait time is chosen so that it is much longer than the characteristic emission time $1/\Gamma = 91~\textrm{ns}$ but much shorter than the time scale of non-radiative processes in the system. As a result the atom is in the driven steady state while fluorescence is being integrated. At low power, $\Omega \ll \Gamma$, the reflection amplitude $r$ as a function of frequency becomes a circle in the parametric IQ-plot, bounded by the values $\textrm{Re}[r] = 1$ for $|\omega-\omega_{03}|\gg \Gamma$ and $\textrm{Re}[r] = 1-2p_0^{\mathrm{th}}$ at $\omega = \omega_{03}$. As stronger driving saturates the $|0\rangle$-$|3\rangle$ transition, the circle deforms into an ellipse and progressively shrinks into a point at $r=1$ for $\Omega \gg \Gamma$. This good agreement between spectroscopy data and the Eq.~(\ref{equ1}) model indicates that transition $|0\rangle$-$|3\rangle$ is coupled to itinerant radiation with a nearly unit efficiency, similarly to previous works on artificial two-level atoms. 

The characterization presented in Fig.~\ref{fig2}a,b provides absolute calibration of the measurement records in terms of the ground state population $p_0$. This is in regard with standard superconducting qubit readout based on the dispersive interaction with a far-detuned cavity, where the measurement records are only known up to a scaling factor to the qubit population and have to be independently calibrated. In order to account for small thermal fluctuations over the course of the experiment, we repeated the previous calibration over 5 days. We find an average thermal equilibrium value of $p_0^\textrm{th} = 0.76\pm0.03$, which corresponds to the atom being at thermal equilibrium with the effective temperature $T = 45\pm5~\textrm{mK}$, which is indeed commonly encountered in superconducting qubits. In what follows, we operate the readout at the optimal power/frequency settings (Fig.~\ref{fig2}a -- black star) where the reflection amplitude becomes a real positive number between zero and unity, with $p_0 = 1-r$.

The protocol for time-domain qubit manipulations matches that of traditional circuit quantum electrodynamics. For example, applying a drive at the qubit frequency $\omega_{01} = 2\pi \times 1.15~\textrm{GHz}$ prior to the readout pulse results in Rabi oscillations of $p_0$ as a function of drive duration (Fig.~\ref{fig2}c - left panel). Operating the circuit slightly off the sweet-spot at $\phi_{\textrm{ext}}/2\pi = 0.507$, i.e. weakly breaking the parity selection rule, we could produce Rabi oscillations between states $|0\rangle$ and $|2\rangle$ by compensating the vanishing transition dipole with a strong drive at frequency $\omega_{02} = 2\pi \times3.88~\textrm{GHz}$~\cite{supp}. Importantly, the non-zero decay rates of even transitions resulting from breaking the  parity selection rule remain several orders of magnitude smaller than other decoherence processes~\cite{supp}. The reflection coefficient measured during Rabi oscillations between $|0\rangle$ and $|1\rangle$ or $|2\rangle$ yields equilibrium ratios $p_1^{\textrm{th}}/p_0^{\textrm{th}} = 0.329$ and $p_2^{\textrm{th}}/p_0^{\textrm{th}} = 0.007$ independently of the absolute calibration of $p_0^{\textrm{th}}$. Assuming $p_0^{\textrm{th}} + p_1^{\textrm{th}} + p_2^{\textrm{th}} = 1$, we obtain the equilibrium value of $p_0^{\textrm{th}} = 0.75$, in close agreement with the spectroscopic calibration. We checked that the qubit can be initialized with $p_0 \approx 0.9$ by strongly driving the $|1\rangle$-$|3\rangle$ transition. During such a drive, entropy is removed from the on-chip circuit via spontaneous emission of photons $|3\rangle \rightarrow|0\rangle$ into the measurement line~\cite{supp}. However, the value $p_0^{\textrm{th}}$ turns out to be sufficiently high for the purpose of our time-domain experiments and hence we performed them all starting from thermal equilibrium. 
 
\begin{figure*}[ht]
	\centering
	\includegraphics[width=.7\linewidth]{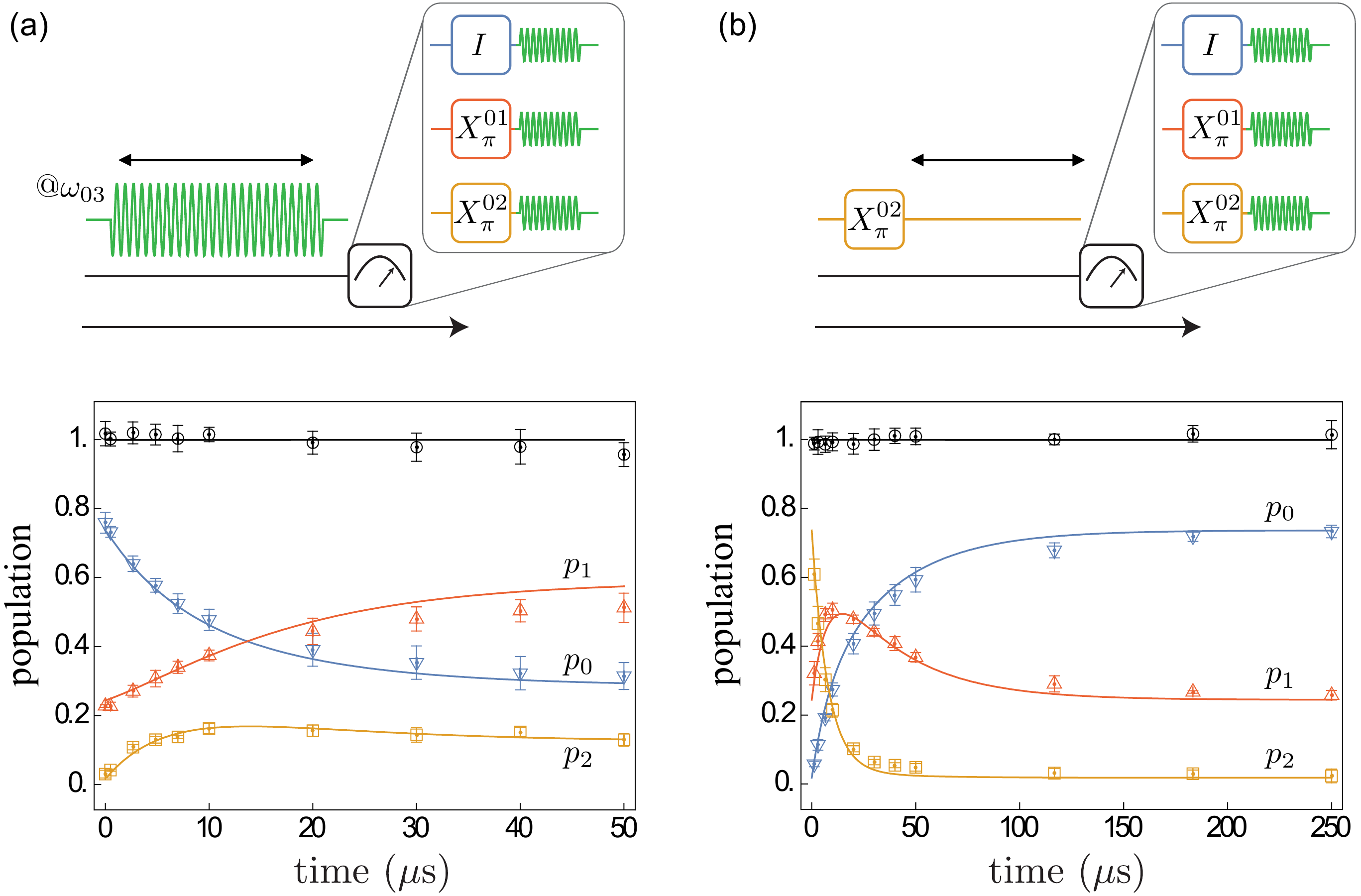}
	\caption {\textbf{Transient dynamics of fluorescence.} (a) Time-domain evolution of populations $p_0$ (down blue triangles), $p_1$ (up red triangles), and $p_2$ (yellow squares) induced by cycling of the readout transition. Note that $p_0$ decays approximately exponentially in time with a characteristic time scale $\tau_{cyc} = 9.6~\mu s$, after which the population is transfered from state $|0\rangle$ to states $|1\rangle$ and $|2\rangle$. The black circles indicate $p_0+p_1+p_2$ and the error bars represent the standard deviation coming from the repetition of the measurements over two days.
		(b) Free evolution of populations starting predominantly from state $|2\rangle$. The data in both (a) and (b) is adequately explained by an optical pumping model (solid lines) (see text) where the dominant error mechanism is a non-radiative decay due to dielectric loss.}
	\label{fig3}
\end{figure*}

The energy relaxation signal following the $0-1$ swap is exponential with a characteristic decay time $T_1 = 52~\mu\textrm{s}$, which is nearly three orders of magnitude longer than the radiative decay time of the $3$-state (Fig.~\ref{fig2}c - right panel). Yet, given the highly suppressed port coupling at the qubit frequency, such value of $T_1$ is still too short to be accounted for by radiative decay. In fact, the qubit decay is likely due to the dielectric loss in the circuit's antenna capacitance with an effective quality factor $Q_{\textrm{diel}} = 5\times 10^5$. The spin-echo signal decay is also exponential with a characteristic coherence time  $T_2 = 52~\mu\textrm{s}$. The additional dephasing $\Gamma_\varphi=1/T_2-1/2T_1$ is likely due to measurement induced dephasing caused by thermal photons at readout frequency. In fact, the rate of thermal jumps $|0\rangle \rightarrow |3\rangle$ can be linked to the effective photon temperature $T$ at the microwave port via the detailed balance equation $\Gamma_\varphi=\Gamma \exp (-\hbar\omega_{03}/k_B T)$. Plugging in the same value $T = 45\pm5~\textrm{mK}$ corresponding to the equilibrium ground state population leads to a close agreement with the measured dephasing rate. Such an error process could be exponentially suppressed in the future by a moderate improvement of circuit thermalization and increasing the readout transition frequency.

Finally, we explore the quantum non-demolition (QND) character of conditional fluorescence measurement. An elementary QND test consists of measuring the atom population dynamics during readout. To do so we drive the $|0\rangle$-$|3\rangle$ transition for a duration $\tau \gg 1/\Gamma$, pause for a brief period for state $|3\rangle$ to relax, swap the state of interest with state $|0\rangle$, and perform the readout (Fig.~\ref{fig3}a - top panel). As a result, we acquire the population transients $p_0(t)$, $p_1(t)$, and $p_2(t)$ of states $|0\rangle$, $|1\rangle$, and $|3\rangle$, respectively, during the readout. The fidelity of this population measurement is limited by the swap gate fidelities, which is above 99\% for $|0\rangle-|1\rangle$ and above 90\% for $|0\rangle-|2\rangle$~\cite{supp}. The $p_0(t)$-data indicates a gradual leakage of population outside the cycling manifold $\{|0\rangle, |3\rangle\}$ on a characteristic time scale $\tau_{cyc} = 9.6~
\mu\textrm{s}$. Importantly, the leakage involves only states $|1\rangle$ and $|2\rangle$ as the data satisfies $p_0 + p_1 + p_2 =1$ within the measurement uncertainty (Fig.~\ref{fig3}a - bottom panel). The average number of fluorescence cycles $N_{cyc} = \Gamma\times \tau_{cyc} \approx 105$ quantifies the deviation of our readout from ideal QND behavior. 

The finite fluorescence lifetime is expected in presence of non-radiative transitions outside the cycling manifold. Additional information about energy relaxation in our circuit was obtained by measuring populations of states $|0\rangle$,  $|1\rangle$, and  $|2\rangle$ as a function of time following the $|0\rangle$-$|2\rangle$ swap (Fig.~\ref{fig3}b). Combining the two data sets represented on Fig.~\ref{fig3}a and b, we constructed a relaxation model based on two additional decay mechanisms, dielectric loss and quasiparticle tunneling across the weak junction. Indeed, such model adequately fits the six experimental curves with only three adjustable parameters: the dielectric loss quality factor $Q_{diel} = 5\times 10^5$, previously introduced to explain the $|1\rangle\rightarrow|0\rangle$ decay, the dimensionless quasiparticle density $x_{qp} = 10^{-6}$, and a fudge-factor of $2.5$ in front of the $|3\rangle\rightarrow|2\rangle$ decay rate. The presence of additional decay for this precise transition could be due to a nearly resonant two-level system~\cite{klimov2018fluctuations}.  From this model we conclude that fluorescence lifetime originates from two error processes. A smaller contribution comes from the direct decay $|3\rangle \rightarrow |1\rangle$ with a characteristic time $36~\mu s$ due to quasiparticle tunneling. Such process may break the parity selection rule for $\phi_{\textrm{ext}} \neq 0$, a prediction that has not been tested so far. The dominant error though comes from the decay $|3\rangle\rightarrow|2\rangle$ with a characteristic time of $3.6 ~\mu s$ due to dielectric loss in the fluxonium capacitance~\cite{supp}. Therefore, improving our fabrication procedures will likely increase the fluorescence lifetime  and hence the QND-fidelity. Alternatively, with a modified choice of circuit parameters, one can increase the frequency of $|1\rangle$-$|2\rangle$ transition and use it for cycling, in which case the fluorescence lifetime would be limited by the qubit $T_1$.

In summary, we showed that a single macroscopic degree of freedom -- the superconducting phase-difference -- can combine  a highly coherent qubit and its QND readout via conditional fluorescence, in full analogy with Demhelt's electron shelving scheme. The present experimental setup can be supplemented with quantum-limited linear amplifiers~\cite{bergeal2010phase} or recently developed microwave photon counters~\cite{opremcak2018measurement,kono2018quantum, besse2018single} for a single-shot operation. Besides novel quantum optics applications, eliminating the cavity from cQED can save hardware resources in scaling up quantum processors and constructing networks for modular quantum computing. 
\\

\noindent\textbf{Methods} 
\subsubsection*{Derivation of Eq.~(\ref{equ1})}
A drive at angular frequency $\omega$ near readout frequency induces damped Rabi oscillations of the readout transition with the Hamiltonian
\begin{equation}
H/\hbar=\frac{(\omega-\omega_{03})}{2}(|3\rangle\langle3|-|0\rangle\langle0|)-i\frac{\Omega}{2}(|3\rangle\langle0|-|0\rangle\langle3|)
\end{equation}
and $\Omega=2\sqrt{\Gamma}\alpha_\mathrm{in}$. In all generality one should consider all sources of loss and dephasing such as spontaneous emission in the line, non-radiative decay and extra-dephasing due to flux noise, including transitions from the readout subspace to other states of the atom. However, the readout transition has been designed so that its emission rate in the line overcomes by many orders of magnitude the other decay and dephasing processes. Assuming an environment at zero temperature, the density matrix of the atom $\rho$ evolves according to the following Lindblad equation
\begin{equation}
\dot{\rho}=-\frac{i}{\hbar}[H,\rho]+\Gamma\mathcal{D}[|0\rangle\langle3|](\rho)
\end{equation}
where $\mathcal{D}$ is the Lindblad superoperator. The steady-state of this equation yields to the expectation value
\begin{equation}
\langle|0\rangle\langle3|\rangle = \frac{\Omega}{\Gamma}\frac{\Gamma^2/2-i(\omega-\omega_{03})\Gamma}{\Gamma^2/2+\Omega^2+2(\omega-\omega_{03})^2},
\end{equation}
from which we derive Eq.~(\ref{equ1}). The good agreement between the theoretical model and the experimental data validates the assumptions made in the previous derivation.

\subsubsection*{Reflection coefficient calibration}
The measured reflection coefficient is only known up to a scaling factor and has to be calibrated. In fact, the readout drive undergoes several stages of attenuation before reaching the artificial atom, while the fluorescence signal is amplified before digitization. Calibration is performed as follows. The circuit is flux biased at $\phi_\textrm{ext}=0$ and is probed in reflection between 6.3 and 6.6~GHz. The external flux is chosen so that no circuit transition exist in the frequency band of interest. The acquired signal $s_\textrm{cal}(\omega)$ therefore accounts for the temperature-dependent attenuation and filtering of the lines as well as the whole acquisition setup amplification chain. Once the atom is biased at the wanted flux, the reflection coefficient $r$ is deduced from the experimentally measured signal $s_\textrm{exp}(\omega)$ by
\begin{equation}
    r = \frac{s_\textrm{exp}(\omega)}{s_\textrm{cal}(\omega)}10^{(P_\textrm{cal}-P_\textrm{exp})/20}
\end{equation}
with $P_\textrm{cal/exp}$ the room-temperature power of the readout drive during the calibration/experiment, expressed in dBm.
\subsubsection*{Relaxation model of transient dynamics}
The relaxation model used to reproduce the data represented on Fig.~\ref{fig3} considers three distinct decay mechanisms. A jump from a state $|i\rangle$ to $|j\rangle$ can occur due to radiative decay in the line, dielectric loss, or quasiparticle tunneling in the weak junction, with the rates $\Gamma_{ij}^\textrm{rad},\ \Gamma_{ij}^\textrm{diel},\ \Gamma_{ij}^\textrm{qp}$, respectively. All rates are computed assuming an equal bath temperature $T=50$~mK for all decay mechanisms~\cite{supp}.

We simulate the dynamics of the atom using the Python library QuTiP\cite{johansson2013}. The spectrum of the fluxonium hamiltonian $H_f=4E_C(-i\partial\phi)^2+E_L\phi^2-E_J\cos(\phi-\phi_\textrm{ext})$ is first computed in a Hilbert space of dimension $100\times100$, to ensure accuracy of the computed eigenfrequencies $\omega_i$ and eigenstates $|i\rangle,\ i\leq99$. We then solve numerically the Lindblad master equation in the Fock state basis for a smaller Hilbert space containing the first 10 eigenstates. This takes into account the fact that high-energy states do not contribute to the atom dynamics and allows for a faster computing time. The master equation contains all the jump operators $\{\sqrt{\Gamma_{ij}^\textrm{rad}+\Gamma_{ij}^\textrm{diel}+\Gamma_{ij}^\textrm{qp}}|j\rangle\langle i|\}_{i,j\leq9}$ computed from the diagonalization of $H_f$. The initial state of the atom is the thermal equilibrium state $\rho_\textrm{th}$ for Fig.~\ref{fig3}a and a perfect swap between $|0\rangle$ and $|2\rangle$ on $\rho_\textrm{th}$ for Fig.~\ref{fig3}b.

The Hamiltonian for Fig.~\ref{fig3}a is $H=-i\frac{\Omega}{2}(|3\rangle\langle0|-|0\rangle\langle3|)$ and represents the readout drive, while it is 0 for Fig.~\ref{fig3}b. They correspond to computing the evolution in the frame rotating at all the eigenfrequencies, i.e. applying the unitary $U(t)=\exp(-iH_f t/\hbar)=\exp(-i\sum_j\omega_j|j\rangle\langle j|t)$ to the driven and undriven hamiltonians, respectively. We find that the transient dynamics of fluorescence is best reproduced for $\Omega=0.95\Gamma/\sqrt{2}$. This deviation from the ideal value can be due to small drifts between the measurements of Fig.~\ref{fig2} and~\ref{fig3}.\\

\noindent\textbf{Data Availability.} The data that support the findings of this study are available from the corresponding author upon reasonable request. 

\noindent\textbf{Acknowledgement.} We thank Ray Mencia for providing samples at the initial stages of this work and Nathan Langford, Shimon Kolkowitz, and Benjamin Huard for useful discussions. We acknowledge funding from Sloan Foundation, NSF-PFC at JQI, and ARO-MURI. \\

\noindent\textbf{Author contributions.} H.X. fabricated the device and along with Y.H.L acquired and analyzed the data. Y.H.L. designed the coaxial-to-waveguide adaptor and built the low-temperature measurement setup. L.B.N. contributed to circuit design, room-temperature instrumentation, and to identifying decoherence mechanisms. N.C. performed initial conditional fluorescence experiments and developed procedures for analyzing and modeling the data. V.E.M. managed the project. The authors declare no financial conflict of interest.

\newpage
\bibliography{SuperconductingCircuits.bib}

\providecommand{\noopsort}[1]{}\providecommand{\singleletter}[1]{#1}%
\begin{thebibliography}{34}%
\makeatletter
\providecommand \@ifxundefined [1]{%
 \@ifx{#1\undefined}
}%
\providecommand \@ifnum [1]{%
 \ifnum #1\expandafter \@firstoftwo
 \else \expandafter \@secondoftwo
 \fi
}%
\providecommand \@ifx [1]{%
 \ifx #1\expandafter \@firstoftwo
 \else \expandafter \@secondoftwo
 \fi
}%
\providecommand \natexlab [1]{#1}%
\providecommand \enquote  [1]{``#1''}%
\providecommand \bibnamefont  [1]{#1}%
\providecommand \bibfnamefont [1]{#1}%
\providecommand \citenamefont [1]{#1}%
\providecommand \href@noop [0]{\@secondoftwo}%
\providecommand \href [0]{\begingroup \@sanitize@url \@href}%
\providecommand \@href[1]{\@@startlink{#1}\@@href}%
\providecommand \@@href[1]{\endgroup#1\@@endlink}%
\providecommand \@sanitize@url [0]{\catcode `\\12\catcode `\$12\catcode
  `\&12\catcode `\#12\catcode `\^12\catcode `\_12\catcode `\%12\relax}%
\providecommand \@@startlink[1]{}%
\providecommand \@@endlink[0]{}%
\providecommand \url  [0]{\begingroup\@sanitize@url \@url }%
\providecommand \@url [1]{\endgroup\@href {#1}{\urlprefix }}%
\providecommand \urlprefix  [0]{URL }%
\providecommand \Eprint [0]{\href }%
\providecommand \doibase [0]{http://dx.doi.org/}%
\providecommand \selectlanguage [0]{\@gobble}%
\providecommand \bibinfo  [0]{\@secondoftwo}%
\providecommand \bibfield  [0]{\@secondoftwo}%
\providecommand \translation [1]{[#1]}%
\providecommand \BibitemOpen [0]{}%
\providecommand \bibitemStop [0]{}%
\providecommand \bibitemNoStop [0]{.\EOS\space}%
\providecommand \EOS [0]{\spacefactor3000\relax}%
\providecommand \BibitemShut  [1]{\csname bibitem#1\endcsname}%
\let\auto@bib@innerbib\@empty
\bibitem [{\citenamefont {Kimble}(2008)}]{kimble2008quantum}%
  \BibitemOpen
  \bibfield  {author} {\bibinfo {author} {\bibfnamefont {H~Jeff}\ \bibnamefont
  {Kimble}},\ }\bibfield  {title} {\enquote {\bibinfo {title} {The quantum
  internet},}\ }\href@noop {} {\bibfield  {journal} {\bibinfo  {journal}
  {Nature}\ }\textbf {\bibinfo {volume} {453}},\ \bibinfo {pages} {1023--1030}
  (\bibinfo {year} {2008})}\BibitemShut {NoStop}%
\bibitem [{\citenamefont {Haroche}\ and\ \citenamefont
  {Raimond}(2006)}]{haroche2006exploring}%
  \BibitemOpen
  \bibfield  {author} {\bibinfo {author} {\bibfnamefont {Serge}\ \bibnamefont
  {Haroche}}\ and\ \bibinfo {author} {\bibfnamefont {Jean-Michel}\ \bibnamefont
  {Raimond}},\ }\href@noop {} {\emph {\bibinfo {title} {Exploring the quantum:
  atoms, cavities, and photons}}}\ (\bibinfo  {publisher} {Oxford university
  press},\ \bibinfo {year} {2006})\BibitemShut {NoStop}%
\bibitem [{\citenamefont {Wallraff}\ \emph {et~al.}(2004)\citenamefont
  {Wallraff}, \citenamefont {Schuster}, \citenamefont {Blais}, \citenamefont
  {Frunzio}, \citenamefont {Huang}, \citenamefont {Majer}, \citenamefont
  {Kumar}, \citenamefont {Girvin},\ and\ \citenamefont
  {Schoelkopf}}]{wallraff2004strong}%
  \BibitemOpen
  \bibfield  {author} {\bibinfo {author} {\bibfnamefont {Andreas}\ \bibnamefont
  {Wallraff}}, \bibinfo {author} {\bibfnamefont {David~I}\ \bibnamefont
  {Schuster}}, \bibinfo {author} {\bibfnamefont {Alexandre}\ \bibnamefont
  {Blais}}, \bibinfo {author} {\bibfnamefont {L}~\bibnamefont {Frunzio}},
  \bibinfo {author} {\bibfnamefont {R-S}\ \bibnamefont {Huang}}, \bibinfo
  {author} {\bibfnamefont {J}~\bibnamefont {Majer}}, \bibinfo {author}
  {\bibfnamefont {S}~\bibnamefont {Kumar}}, \bibinfo {author} {\bibfnamefont
  {Steven~M}\ \bibnamefont {Girvin}}, \ and\ \bibinfo {author} {\bibfnamefont
  {Robert~J}\ \bibnamefont {Schoelkopf}},\ }\bibfield  {title} {\enquote
  {\bibinfo {title} {Strong coupling of a single photon to a superconducting
  qubit using circuit quantum electrodynamics},}\ }\href@noop {} {\bibfield
  {journal} {\bibinfo  {journal} {Nature}\ }\textbf {\bibinfo {volume} {431}},\
  \bibinfo {pages} {162--167} (\bibinfo {year} {2004})}\BibitemShut {NoStop}%
\bibitem [{\citenamefont {Haroche}\ \emph {et~al.}(2020)\citenamefont
  {Haroche}, \citenamefont {Brune},\ and\ \citenamefont
  {Raimond}}]{haroche2020cavity}%
  \BibitemOpen
  \bibfield  {author} {\bibinfo {author} {\bibfnamefont {S}~\bibnamefont
  {Haroche}}, \bibinfo {author} {\bibfnamefont {M}~\bibnamefont {Brune}}, \
  and\ \bibinfo {author} {\bibfnamefont {JM}~\bibnamefont {Raimond}},\
  }\bibfield  {title} {\enquote {\bibinfo {title} {From cavity to circuit
  quantum electrodynamics},}\ }\href@noop {} {\bibfield  {journal} {\bibinfo
  {journal} {Nature Physics}\ ,\ \bibinfo {pages} {1--4}} (\bibinfo {year}
  {2020})}\BibitemShut {NoStop}%
\bibitem [{\citenamefont {Blais}\ \emph {et~al.}(2020)\citenamefont {Blais},
  \citenamefont {Girvin},\ and\ \citenamefont {Oliver}}]{blais2020quantum}%
  \BibitemOpen
  \bibfield  {author} {\bibinfo {author} {\bibfnamefont {Alexandre}\
  \bibnamefont {Blais}}, \bibinfo {author} {\bibfnamefont {Steven~M}\
  \bibnamefont {Girvin}}, \ and\ \bibinfo {author} {\bibfnamefont {William~D}\
  \bibnamefont {Oliver}},\ }\bibfield  {title} {\enquote {\bibinfo {title}
  {Quantum information processing and quantum optics with circuit quantum
  electrodynamics},}\ }\href@noop {} {\bibfield  {journal} {\bibinfo  {journal}
  {Nature Physics}\ ,\ \bibinfo {pages} {1--10}} (\bibinfo {year}
  {2020})}\BibitemShut {NoStop}%
\bibitem [{\citenamefont {Ludlow}\ \emph {et~al.}(2015)\citenamefont {Ludlow},
  \citenamefont {Boyd}, \citenamefont {Ye}, \citenamefont {Peik},\ and\
  \citenamefont {Schmidt}}]{ludlow2015optical}%
  \BibitemOpen
  \bibfield  {author} {\bibinfo {author} {\bibfnamefont {Andrew~D}\
  \bibnamefont {Ludlow}}, \bibinfo {author} {\bibfnamefont {Martin~M}\
  \bibnamefont {Boyd}}, \bibinfo {author} {\bibfnamefont {Jun}\ \bibnamefont
  {Ye}}, \bibinfo {author} {\bibfnamefont {Ekkehard}\ \bibnamefont {Peik}}, \
  and\ \bibinfo {author} {\bibfnamefont {Piet~O}\ \bibnamefont {Schmidt}},\
  }\bibfield  {title} {\enquote {\bibinfo {title} {Optical atomic clocks},}\
  }\href@noop {} {\bibfield  {journal} {\bibinfo  {journal} {Rev. Mod. Phys.}\
  }\textbf {\bibinfo {volume} {87}},\ \bibinfo {pages} {637} (\bibinfo {year}
  {2015})}\BibitemShut {NoStop}%
\bibitem [{\citenamefont {Nagourney}\ \emph {et~al.}(1986)\citenamefont
  {Nagourney}, \citenamefont {Sandberg},\ and\ \citenamefont
  {Dehmelt}}]{nagourney1986shelved}%
  \BibitemOpen
  \bibfield  {author} {\bibinfo {author} {\bibfnamefont {Warren}\ \bibnamefont
  {Nagourney}}, \bibinfo {author} {\bibfnamefont {Jon}\ \bibnamefont
  {Sandberg}}, \ and\ \bibinfo {author} {\bibfnamefont {Hans}\ \bibnamefont
  {Dehmelt}},\ }\bibfield  {title} {\enquote {\bibinfo {title} {Shelved optical
  electron amplifier: Observation of quantum jumps},}\ }\href@noop {}
  {\bibfield  {journal} {\bibinfo  {journal} {Physical Review Letters}\
  }\textbf {\bibinfo {volume} {56}},\ \bibinfo {pages} {2797} (\bibinfo {year}
  {1986})}\BibitemShut {NoStop}%
\bibitem [{\citenamefont {Bruzewicz}\ \emph {et~al.}(2019)\citenamefont
  {Bruzewicz}, \citenamefont {Chiaverini}, \citenamefont {McConnell},\ and\
  \citenamefont {Sage}}]{bruzewicz2019trapped}%
  \BibitemOpen
  \bibfield  {author} {\bibinfo {author} {\bibfnamefont {Colin~D}\ \bibnamefont
  {Bruzewicz}}, \bibinfo {author} {\bibfnamefont {John}\ \bibnamefont
  {Chiaverini}}, \bibinfo {author} {\bibfnamefont {Robert}\ \bibnamefont
  {McConnell}}, \ and\ \bibinfo {author} {\bibfnamefont {Jeremy~M}\
  \bibnamefont {Sage}},\ }\bibfield  {title} {\enquote {\bibinfo {title}
  {Trapped-ion quantum computing: Progress and challenges},}\ }\href@noop {}
  {\bibfield  {journal} {\bibinfo  {journal} {Applied Physics Reviews}\
  }\textbf {\bibinfo {volume} {6}},\ \bibinfo {pages} {021314} (\bibinfo {year}
  {2019})}\BibitemShut {NoStop}%
\bibitem [{\citenamefont {Robledo}\ \emph {et~al.}(2011)\citenamefont
  {Robledo}, \citenamefont {Childress}, \citenamefont {Bernien}, \citenamefont
  {Hensen}, \citenamefont {Alkemade},\ and\ \citenamefont
  {Hanson}}]{robledo2011high}%
  \BibitemOpen
  \bibfield  {author} {\bibinfo {author} {\bibfnamefont {Lucio}\ \bibnamefont
  {Robledo}}, \bibinfo {author} {\bibfnamefont {Lilian}\ \bibnamefont
  {Childress}}, \bibinfo {author} {\bibfnamefont {Hannes}\ \bibnamefont
  {Bernien}}, \bibinfo {author} {\bibfnamefont {Bas}\ \bibnamefont {Hensen}},
  \bibinfo {author} {\bibfnamefont {Paul~FA}\ \bibnamefont {Alkemade}}, \ and\
  \bibinfo {author} {\bibfnamefont {Ronald}\ \bibnamefont {Hanson}},\
  }\bibfield  {title} {\enquote {\bibinfo {title} {High-fidelity projective
  read-out of a solid-state spin quantum register},}\ }\href@noop {} {\bibfield
   {journal} {\bibinfo  {journal} {Nature}\ }\textbf {\bibinfo {volume}
  {477}},\ \bibinfo {pages} {574--578} (\bibinfo {year} {2011})}\BibitemShut
  {NoStop}%
\bibitem [{\citenamefont {Sukachev}\ \emph {et~al.}(2017)\citenamefont
  {Sukachev}, \citenamefont {Sipahigil}, \citenamefont {Nguyen}, \citenamefont
  {Bhaskar}, \citenamefont {Evans}, \citenamefont {Jelezko},\ and\
  \citenamefont {Lukin}}]{sukachev2017silicon}%
  \BibitemOpen
  \bibfield  {author} {\bibinfo {author} {\bibfnamefont {Denis~D}\ \bibnamefont
  {Sukachev}}, \bibinfo {author} {\bibfnamefont {Alp}\ \bibnamefont
  {Sipahigil}}, \bibinfo {author} {\bibfnamefont {Christian~T}\ \bibnamefont
  {Nguyen}}, \bibinfo {author} {\bibfnamefont {Mihir~K}\ \bibnamefont
  {Bhaskar}}, \bibinfo {author} {\bibfnamefont {Ruffin~E}\ \bibnamefont
  {Evans}}, \bibinfo {author} {\bibfnamefont {Fedor}\ \bibnamefont {Jelezko}},
  \ and\ \bibinfo {author} {\bibfnamefont {Mikhail~D}\ \bibnamefont {Lukin}},\
  }\bibfield  {title} {\enquote {\bibinfo {title} {Silicon-vacancy spin qubit
  in diamond: a quantum memory exceeding 10 ms with single-shot state
  readout},}\ }\href@noop {} {\bibfield  {journal} {\bibinfo  {journal}
  {Physical review letters}\ }\textbf {\bibinfo {volume} {119}},\ \bibinfo
  {pages} {223602} (\bibinfo {year} {2017})}\BibitemShut {NoStop}%
\bibitem [{\citenamefont {Vamivakas}\ \emph {et~al.}(2009)\citenamefont
  {Vamivakas}, \citenamefont {Zhao}, \citenamefont {Lu},\ and\ \citenamefont
  {Atat{\"u}re}}]{vamivakas2009spin}%
  \BibitemOpen
  \bibfield  {author} {\bibinfo {author} {\bibfnamefont {A~Nick}\ \bibnamefont
  {Vamivakas}}, \bibinfo {author} {\bibfnamefont {Yong}\ \bibnamefont {Zhao}},
  \bibinfo {author} {\bibfnamefont {Chao-Yang}\ \bibnamefont {Lu}}, \ and\
  \bibinfo {author} {\bibfnamefont {Mete}\ \bibnamefont {Atat{\"u}re}},\
  }\bibfield  {title} {\enquote {\bibinfo {title} {Spin-resolved quantum-dot
  resonance fluorescence},}\ }\href@noop {} {\bibfield  {journal} {\bibinfo
  {journal} {Nature Physics}\ }\textbf {\bibinfo {volume} {5}},\ \bibinfo
  {pages} {198--202} (\bibinfo {year} {2009})}\BibitemShut {NoStop}%
\bibitem [{\citenamefont {Manucharyan}\ \emph {et~al.}(2009)\citenamefont
  {Manucharyan}, \citenamefont {Koch}, \citenamefont {Glazman},\ and\
  \citenamefont {Devoret}}]{Manucharyan113}%
  \BibitemOpen
  \bibfield  {author} {\bibinfo {author} {\bibfnamefont {Vladimir~E.}\
  \bibnamefont {Manucharyan}}, \bibinfo {author} {\bibfnamefont {Jens}\
  \bibnamefont {Koch}}, \bibinfo {author} {\bibfnamefont {Leonid~I.}\
  \bibnamefont {Glazman}}, \ and\ \bibinfo {author} {\bibfnamefont {Michel~H.}\
  \bibnamefont {Devoret}},\ }\bibfield  {title} {\enquote {\bibinfo {title}
  {Fluxonium: Single cooper-pair circuit free of charge offsets},}\ }\href
  {\doibase 10.1126/science.1175552} {\bibfield  {journal} {\bibinfo  {journal}
  {Science}\ }\textbf {\bibinfo {volume} {326}},\ \bibinfo {pages} {113--116}
  (\bibinfo {year} {2009})}\BibitemShut {NoStop}%
\bibitem [{\citenamefont {Astafiev}\ \emph {et~al.}(2010)\citenamefont
  {Astafiev}, \citenamefont {Zagoskin}, \citenamefont {Abdumalikov},
  \citenamefont {Pashkin}, \citenamefont {Yamamoto}, \citenamefont {Inomata},
  \citenamefont {Nakamura},\ and\ \citenamefont
  {Tsai}}]{astafiev2010resonance}%
  \BibitemOpen
  \bibfield  {author} {\bibinfo {author} {\bibfnamefont {O}~\bibnamefont
  {Astafiev}}, \bibinfo {author} {\bibfnamefont {Alexandre~M}\ \bibnamefont
  {Zagoskin}}, \bibinfo {author} {\bibfnamefont {AA}~\bibnamefont
  {Abdumalikov}}, \bibinfo {author} {\bibfnamefont {Yu~A}\ \bibnamefont
  {Pashkin}}, \bibinfo {author} {\bibfnamefont {T}~\bibnamefont {Yamamoto}},
  \bibinfo {author} {\bibfnamefont {K}~\bibnamefont {Inomata}}, \bibinfo
  {author} {\bibfnamefont {Y}~\bibnamefont {Nakamura}}, \ and\ \bibinfo
  {author} {\bibfnamefont {Jaw~Shen}\ \bibnamefont {Tsai}},\ }\bibfield
  {title} {\enquote {\bibinfo {title} {Resonance fluorescence of a single
  artificial atom},}\ }\href@noop {} {\bibfield  {journal} {\bibinfo  {journal}
  {Science}\ }\textbf {\bibinfo {volume} {327}},\ \bibinfo {pages} {840--843}
  (\bibinfo {year} {2010})}\BibitemShut {NoStop}%
\bibitem [{\citenamefont {Braginsky}\ and\ \citenamefont
  {Khalili}(1996)}]{braginsky1996quantum}%
  \BibitemOpen
  \bibfield  {author} {\bibinfo {author} {\bibfnamefont {Vladimir~B}\
  \bibnamefont {Braginsky}}\ and\ \bibinfo {author} {\bibfnamefont {F~Ya}\
  \bibnamefont {Khalili}},\ }\bibfield  {title} {\enquote {\bibinfo {title}
  {Quantum nondemolition measurements: the route from toys to tools},}\
  }\href@noop {} {\bibfield  {journal} {\bibinfo  {journal} {Reviews of Modern
  Physics}\ }\textbf {\bibinfo {volume} {68}},\ \bibinfo {pages} {1} (\bibinfo
  {year} {1996})}\BibitemShut {NoStop}%
\bibitem [{\citenamefont {Hoi}\ \emph {et~al.}(2011)\citenamefont {Hoi},
  \citenamefont {Wilson}, \citenamefont {Johansson}, \citenamefont {Palomaki},
  \citenamefont {Peropadre},\ and\ \citenamefont
  {Delsing}}]{hoi2011demonstration}%
  \BibitemOpen
  \bibfield  {author} {\bibinfo {author} {\bibfnamefont {Io-Chun}\ \bibnamefont
  {Hoi}}, \bibinfo {author} {\bibfnamefont {C.~M.}\ \bibnamefont {Wilson}},
  \bibinfo {author} {\bibfnamefont {G{\"o}ran}\ \bibnamefont {Johansson}},
  \bibinfo {author} {\bibfnamefont {Tauno}\ \bibnamefont {Palomaki}}, \bibinfo
  {author} {\bibfnamefont {Borja}\ \bibnamefont {Peropadre}}, \ and\ \bibinfo
  {author} {\bibfnamefont {Per}\ \bibnamefont {Delsing}},\ }\bibfield  {title}
  {\enquote {\bibinfo {title} {Demonstration of a single-photon router in the
  microwave regime},}\ }\href@noop {} {\bibfield  {journal} {\bibinfo
  {journal} {Physical Review Letters}\ }\textbf {\bibinfo {volume} {107}},\
  \bibinfo {pages} {073601} (\bibinfo {year} {2011})}\BibitemShut {NoStop}%
\bibitem [{\citenamefont {Van~Loo}\ \emph {et~al.}(2013)\citenamefont
  {Van~Loo}, \citenamefont {Fedorov}, \citenamefont {Lalumiere}, \citenamefont
  {Sanders}, \citenamefont {Blais},\ and\ \citenamefont
  {Wallraff}}]{van2013photon}%
  \BibitemOpen
  \bibfield  {author} {\bibinfo {author} {\bibfnamefont {Arjan~F}\ \bibnamefont
  {Van~Loo}}, \bibinfo {author} {\bibfnamefont {Arkady}\ \bibnamefont
  {Fedorov}}, \bibinfo {author} {\bibfnamefont {Kevin}\ \bibnamefont
  {Lalumiere}}, \bibinfo {author} {\bibfnamefont {Barry~C}\ \bibnamefont
  {Sanders}}, \bibinfo {author} {\bibfnamefont {Alexandre}\ \bibnamefont
  {Blais}}, \ and\ \bibinfo {author} {\bibfnamefont {Andreas}\ \bibnamefont
  {Wallraff}},\ }\bibfield  {title} {\enquote {\bibinfo {title}
  {Photon-mediated interactions between distant artificial atoms},}\
  }\href@noop {} {\bibfield  {journal} {\bibinfo  {journal} {Science}\ }\textbf
  {\bibinfo {volume} {342}},\ \bibinfo {pages} {1494--1496} (\bibinfo {year}
  {2013})}\BibitemShut {NoStop}%
\bibitem [{\citenamefont {Mirhosseini}\ \emph {et~al.}(2019)\citenamefont
  {Mirhosseini}, \citenamefont {Kim}, \citenamefont {Zhang}, \citenamefont
  {Sipahigil}, \citenamefont {Dieterle}, \citenamefont {Keller}, \citenamefont
  {Asenjo-Garcia}, \citenamefont {Chang},\ and\ \citenamefont
  {Painter}}]{mirhosseini2019}%
  \BibitemOpen
  \bibfield  {author} {\bibinfo {author} {\bibfnamefont {Mohammad}\
  \bibnamefont {Mirhosseini}}, \bibinfo {author} {\bibfnamefont {Eunjong}\
  \bibnamefont {Kim}}, \bibinfo {author} {\bibfnamefont {Xueyue}\ \bibnamefont
  {Zhang}}, \bibinfo {author} {\bibfnamefont {Alp}\ \bibnamefont {Sipahigil}},
  \bibinfo {author} {\bibfnamefont {Paul~B.}\ \bibnamefont {Dieterle}},
  \bibinfo {author} {\bibfnamefont {Andrew~J.}\ \bibnamefont {Keller}},
  \bibinfo {author} {\bibfnamefont {Ana}\ \bibnamefont {Asenjo-Garcia}},
  \bibinfo {author} {\bibfnamefont {Darrick~E.}\ \bibnamefont {Chang}}, \ and\
  \bibinfo {author} {\bibfnamefont {Oskar}\ \bibnamefont {Painter}},\
  }\bibfield  {title} {\enquote {\bibinfo {title} {{Cavity quantum
  electrodynamics with atom-like mirrors}},}\ }\href {\doibase
  10.1038/s41586-019-1196-1} {\bibfield  {journal} {\bibinfo  {journal}
  {Nature}\ } (\bibinfo {year} {2019}),\ 10.1038/s41586-019-1196-1}\BibitemShut
  {NoStop}%
\bibitem [{\citenamefont {Kannan}\ \emph {et~al.}(2020)\citenamefont {Kannan},
  \citenamefont {Campbell}, \citenamefont {Vasconcelos}, \citenamefont {Winik},
  \citenamefont {Kim}, \citenamefont {Kjaergaard}, \citenamefont {Krantz},
  \citenamefont {Melville}, \citenamefont {Niedzielski}, \citenamefont {Yoder},
  \citenamefont {Orlando}, \citenamefont {Gustavsson},\ and\ \citenamefont
  {Oliver}}]{kannan2020}%
  \BibitemOpen
  \bibfield  {author} {\bibinfo {author} {\bibfnamefont {Bharath}\ \bibnamefont
  {Kannan}}, \bibinfo {author} {\bibfnamefont {Daniel}\ \bibnamefont
  {Campbell}}, \bibinfo {author} {\bibfnamefont {Francisca}\ \bibnamefont
  {Vasconcelos}}, \bibinfo {author} {\bibfnamefont {Roni}\ \bibnamefont
  {Winik}}, \bibinfo {author} {\bibfnamefont {David}\ \bibnamefont {Kim}},
  \bibinfo {author} {\bibfnamefont {Morten}\ \bibnamefont {Kjaergaard}},
  \bibinfo {author} {\bibfnamefont {Philip}\ \bibnamefont {Krantz}}, \bibinfo
  {author} {\bibfnamefont {Alexander}\ \bibnamefont {Melville}}, \bibinfo
  {author} {\bibfnamefont {Bethany~M.}\ \bibnamefont {Niedzielski}}, \bibinfo
  {author} {\bibfnamefont {Jonilyn}\ \bibnamefont {Yoder}}, \bibinfo {author}
  {\bibfnamefont {Terry~P.}\ \bibnamefont {Orlando}}, \bibinfo {author}
  {\bibfnamefont {Simon}\ \bibnamefont {Gustavsson}}, \ and\ \bibinfo {author}
  {\bibfnamefont {William~D.}\ \bibnamefont {Oliver}},\ }\bibfield  {title}
  {\enquote {\bibinfo {title} {{Generating Spatially Entangled Itinerant
  Photons with Waveguide Quantum Electrodynamics}},}\ }\href
  {http://arxiv.org/abs/2003.07300} {\  (\bibinfo {year} {2020})},\ \Eprint
  {http://arxiv.org/abs/2003.07300} {arXiv:2003.07300} \BibitemShut {NoStop}%
\bibitem [{\citenamefont {Houck}\ \emph {et~al.}(2007)\citenamefont {Houck},
  \citenamefont {Schuster}, \citenamefont {Gambetta}, \citenamefont {Schreier},
  \citenamefont {Johnson}, \citenamefont {Chow}, \citenamefont {Frunzio},
  \citenamefont {Majer}, \citenamefont {Devoret}, \citenamefont {Girvin} \emph
  {et~al.}}]{houck2007generating}%
  \BibitemOpen
  \bibfield  {author} {\bibinfo {author} {\bibfnamefont {AA}~\bibnamefont
  {Houck}}, \bibinfo {author} {\bibfnamefont {DI}~\bibnamefont {Schuster}},
  \bibinfo {author} {\bibfnamefont {JM}~\bibnamefont {Gambetta}}, \bibinfo
  {author} {\bibfnamefont {JA}~\bibnamefont {Schreier}}, \bibinfo {author}
  {\bibfnamefont {BR}~\bibnamefont {Johnson}}, \bibinfo {author} {\bibfnamefont
  {JM}~\bibnamefont {Chow}}, \bibinfo {author} {\bibfnamefont {L}~\bibnamefont
  {Frunzio}}, \bibinfo {author} {\bibfnamefont {J}~\bibnamefont {Majer}},
  \bibinfo {author} {\bibfnamefont {MH}~\bibnamefont {Devoret}}, \bibinfo
  {author} {\bibfnamefont {SM}~\bibnamefont {Girvin}},  \emph {et~al.},\
  }\bibfield  {title} {\enquote {\bibinfo {title} {Generating single microwave
  photons in a circuit},}\ }\href@noop {} {\bibfield  {journal} {\bibinfo
  {journal} {Nature}\ }\textbf {\bibinfo {volume} {449}},\ \bibinfo {pages}
  {328--331} (\bibinfo {year} {2007})}\BibitemShut {NoStop}%
\bibitem [{\citenamefont {Campagne-Ibarcq}\ \emph {et~al.}(2014)\citenamefont
  {Campagne-Ibarcq}, \citenamefont {Bretheau}, \citenamefont {Flurin},
  \citenamefont {Auff{\`e}ves}, \citenamefont {Mallet},\ and\ \citenamefont
  {Huard}}]{campagne2014observing}%
  \BibitemOpen
  \bibfield  {author} {\bibinfo {author} {\bibfnamefont {Philippe}\
  \bibnamefont {Campagne-Ibarcq}}, \bibinfo {author} {\bibfnamefont {Landry}\
  \bibnamefont {Bretheau}}, \bibinfo {author} {\bibfnamefont {Emmanuel}\
  \bibnamefont {Flurin}}, \bibinfo {author} {\bibfnamefont {Alexia}\
  \bibnamefont {Auff{\`e}ves}}, \bibinfo {author} {\bibfnamefont
  {Fran{\c{c}}ois}\ \bibnamefont {Mallet}}, \ and\ \bibinfo {author}
  {\bibfnamefont {Benjamin}\ \bibnamefont {Huard}},\ }\bibfield  {title}
  {\enquote {\bibinfo {title} {Observing interferences between past and future
  quantum states in resonance fluorescence},}\ }\href@noop {} {\bibfield
  {journal} {\bibinfo  {journal} {Physical Review Letters}\ }\textbf {\bibinfo
  {volume} {112}},\ \bibinfo {pages} {180402} (\bibinfo {year}
  {2014})}\BibitemShut {NoStop}%
\bibitem [{\citenamefont {Toyli}\ \emph {et~al.}(2016)\citenamefont {Toyli},
  \citenamefont {Eddins}, \citenamefont {Boutin}, \citenamefont {Puri},
  \citenamefont {Hover}, \citenamefont {Bolkhovsky}, \citenamefont {Oliver},
  \citenamefont {Blais},\ and\ \citenamefont {Siddiqi}}]{toyli2016resonance}%
  \BibitemOpen
  \bibfield  {author} {\bibinfo {author} {\bibfnamefont {DM}~\bibnamefont
  {Toyli}}, \bibinfo {author} {\bibfnamefont {AW}~\bibnamefont {Eddins}},
  \bibinfo {author} {\bibfnamefont {S}~\bibnamefont {Boutin}}, \bibinfo
  {author} {\bibfnamefont {S}~\bibnamefont {Puri}}, \bibinfo {author}
  {\bibfnamefont {D}~\bibnamefont {Hover}}, \bibinfo {author} {\bibfnamefont
  {V}~\bibnamefont {Bolkhovsky}}, \bibinfo {author} {\bibfnamefont
  {WD}~\bibnamefont {Oliver}}, \bibinfo {author} {\bibfnamefont
  {A}~\bibnamefont {Blais}}, \ and\ \bibinfo {author} {\bibfnamefont
  {I}~\bibnamefont {Siddiqi}},\ }\bibfield  {title} {\enquote {\bibinfo {title}
  {Resonance fluorescence from an artificial atom in squeezed vacuum},}\
  }\href@noop {} {\bibfield  {journal} {\bibinfo  {journal} {Physical Review
  X}\ }\textbf {\bibinfo {volume} {6}},\ \bibinfo {pages} {031004} (\bibinfo
  {year} {2016})}\BibitemShut {NoStop}%
\bibitem [{\citenamefont {Cottet}\ \emph {et~al.}(2017)\citenamefont {Cottet},
  \citenamefont {Jezouin}, \citenamefont {Bretheau}, \citenamefont
  {Campagne-Ibarcq}, \citenamefont {Ficheux}, \citenamefont {Anders},
  \citenamefont {Auff{\`{e}}ves}, \citenamefont {Azouit}, \citenamefont
  {Rouchon},\ and\ \citenamefont {Huard}}]{cottet2017}%
  \BibitemOpen
  \bibfield  {author} {\bibinfo {author} {\bibfnamefont {Nathana{\"{e}}l}\
  \bibnamefont {Cottet}}, \bibinfo {author} {\bibfnamefont {S{\'{e}}bastien}\
  \bibnamefont {Jezouin}}, \bibinfo {author} {\bibfnamefont {Landry}\
  \bibnamefont {Bretheau}}, \bibinfo {author} {\bibfnamefont {Philippe}\
  \bibnamefont {Campagne-Ibarcq}}, \bibinfo {author} {\bibfnamefont {Quentin}\
  \bibnamefont {Ficheux}}, \bibinfo {author} {\bibfnamefont {Janet}\
  \bibnamefont {Anders}}, \bibinfo {author} {\bibfnamefont {Alexia}\
  \bibnamefont {Auff{\`{e}}ves}}, \bibinfo {author} {\bibfnamefont
  {R{\'{e}}mi}\ \bibnamefont {Azouit}}, \bibinfo {author} {\bibfnamefont
  {Pierre}\ \bibnamefont {Rouchon}}, \ and\ \bibinfo {author} {\bibfnamefont
  {Benjamin}\ \bibnamefont {Huard}},\ }\bibfield  {title} {\enquote {\bibinfo
  {title} {{Observing a quantum Maxwell demon at work}},}\ }\href {\doibase
  10.1073/pnas.1704827114} {\bibfield  {journal} {\bibinfo  {journal}
  {Proceedings of the National Academy of Sciences}\ }\textbf {\bibinfo
  {volume} {114}},\ \bibinfo {pages} {7561--7564} (\bibinfo {year} {2017})},\
  \Eprint {http://arxiv.org/abs/1702.05161} {arXiv:1702.05161} \BibitemShut
  {NoStop}%
\bibitem [{\citenamefont {Englert}\ \emph {et~al.}(2010)\citenamefont
  {Englert}, \citenamefont {Mangano}, \citenamefont {Mariantoni}, \citenamefont
  {Gross}, \citenamefont {Siewert},\ and\ \citenamefont
  {Solano}}]{englert2010mesoscopic}%
  \BibitemOpen
  \bibfield  {author} {\bibinfo {author} {\bibfnamefont {Barbara
  Gabriele~Ursula}\ \bibnamefont {Englert}}, \bibinfo {author} {\bibfnamefont
  {G}~\bibnamefont {Mangano}}, \bibinfo {author} {\bibfnamefont
  {M}~\bibnamefont {Mariantoni}}, \bibinfo {author} {\bibfnamefont
  {R}~\bibnamefont {Gross}}, \bibinfo {author} {\bibfnamefont {J}~\bibnamefont
  {Siewert}}, \ and\ \bibinfo {author} {\bibfnamefont {E}~\bibnamefont
  {Solano}},\ }\bibfield  {title} {\enquote {\bibinfo {title} {Mesoscopic
  shelving readout of superconducting qubits in circuit quantum
  electrodynamics},}\ }\href@noop {} {\bibfield  {journal} {\bibinfo  {journal}
  {Physical Review B}\ }\textbf {\bibinfo {volume} {81}},\ \bibinfo {pages}
  {134514} (\bibinfo {year} {2010})}\BibitemShut {NoStop}%
\bibitem [{\citenamefont {Nguyen}\ \emph {et~al.}(2019)\citenamefont {Nguyen},
  \citenamefont {Lin}, \citenamefont {Somoroff}, \citenamefont {Mencia},
  \citenamefont {Grabon},\ and\ \citenamefont {Manucharyan}}]{nguyen2019high}%
  \BibitemOpen
  \bibfield  {author} {\bibinfo {author} {\bibfnamefont {Long~B}\ \bibnamefont
  {Nguyen}}, \bibinfo {author} {\bibfnamefont {Yen-Hsiang}\ \bibnamefont
  {Lin}}, \bibinfo {author} {\bibfnamefont {Aaron}\ \bibnamefont {Somoroff}},
  \bibinfo {author} {\bibfnamefont {Raymond}\ \bibnamefont {Mencia}}, \bibinfo
  {author} {\bibfnamefont {Nicholas}\ \bibnamefont {Grabon}}, \ and\ \bibinfo
  {author} {\bibfnamefont {Vladimir~E}\ \bibnamefont {Manucharyan}},\
  }\bibfield  {title} {\enquote {\bibinfo {title} {High-coherence fluxonium
  qubit},}\ }\href@noop {} {\bibfield  {journal} {\bibinfo  {journal} {Physical
  Review X}\ }\textbf {\bibinfo {volume} {9}},\ \bibinfo {pages} {041041}
  (\bibinfo {year} {2019})}\BibitemShut {NoStop}%
\bibitem [{\citenamefont {Zhang}\ \emph {et~al.}(2020)\citenamefont {Zhang},
  \citenamefont {Chakram}, \citenamefont {Roy}, \citenamefont {Earnest},
  \citenamefont {Lu}, \citenamefont {Huang}, \citenamefont {Weiss},
  \citenamefont {Koch},\ and\ \citenamefont {Schuster}}]{zhang2020universal}%
  \BibitemOpen
  \bibfield  {author} {\bibinfo {author} {\bibfnamefont {Helin}\ \bibnamefont
  {Zhang}}, \bibinfo {author} {\bibfnamefont {Srivatsan}\ \bibnamefont
  {Chakram}}, \bibinfo {author} {\bibfnamefont {Tanay}\ \bibnamefont {Roy}},
  \bibinfo {author} {\bibfnamefont {Nathan}\ \bibnamefont {Earnest}}, \bibinfo
  {author} {\bibfnamefont {Yao}\ \bibnamefont {Lu}}, \bibinfo {author}
  {\bibfnamefont {Ziwen}\ \bibnamefont {Huang}}, \bibinfo {author}
  {\bibfnamefont {Daniel}\ \bibnamefont {Weiss}}, \bibinfo {author}
  {\bibfnamefont {Jens}\ \bibnamefont {Koch}}, \ and\ \bibinfo {author}
  {\bibfnamefont {David~I}\ \bibnamefont {Schuster}},\ }\bibfield  {title}
  {\enquote {\bibinfo {title} {Universal fast flux control of a coherent,
  low-frequency qubit},}\ }\href@noop {} {\bibfield  {journal} {\bibinfo
  {journal} {arXiv preprint arXiv:2002.10653}\ } (\bibinfo {year}
  {2020})}\BibitemShut {NoStop}%
\bibitem [{sup()}]{supp}%
  \BibitemOpen
  \href@noop {} {\emph {\bibinfo {title} {See Supplementary
  Material}}}\BibitemShut {NoStop}%
\bibitem [{\citenamefont {Kou}\ \emph {et~al.}(2018)\citenamefont {Kou},
  \citenamefont {Smith}, \citenamefont {Vool}, \citenamefont {Pop},
  \citenamefont {Sliwa}, \citenamefont {Hatridge}, \citenamefont {Frunzio},\
  and\ \citenamefont {Devoret}}]{kou2018simultaneous}%
  \BibitemOpen
  \bibfield  {author} {\bibinfo {author} {\bibfnamefont {A}~\bibnamefont
  {Kou}}, \bibinfo {author} {\bibfnamefont {W.~C.}\ \bibnamefont {Smith}},
  \bibinfo {author} {\bibfnamefont {U}~\bibnamefont {Vool}}, \bibinfo {author}
  {\bibfnamefont {I.~M.}\ \bibnamefont {Pop}}, \bibinfo {author} {\bibfnamefont
  {K.~M.}\ \bibnamefont {Sliwa}}, \bibinfo {author} {\bibfnamefont
  {M}~\bibnamefont {Hatridge}}, \bibinfo {author} {\bibfnamefont
  {L}~\bibnamefont {Frunzio}}, \ and\ \bibinfo {author} {\bibfnamefont {M.~H.}\
  \bibnamefont {Devoret}},\ }\bibfield  {title} {\enquote {\bibinfo {title}
  {Simultaneous monitoring of fluxonium qubits in a waveguide},}\ }\href@noop
  {} {\bibfield  {journal} {\bibinfo  {journal} {Physical Review Applied}\
  }\textbf {\bibinfo {volume} {9}},\ \bibinfo {pages} {064022} (\bibinfo {year}
  {2018})}\BibitemShut {NoStop}%
\bibitem [{\citenamefont {Gardiner}\ and\ \citenamefont
  {Collett}(1985)}]{gardiner1985}%
  \BibitemOpen
  \bibfield  {author} {\bibinfo {author} {\bibfnamefont {C.~W.}\ \bibnamefont
  {Gardiner}}\ and\ \bibinfo {author} {\bibfnamefont {M.~J.}\ \bibnamefont
  {Collett}},\ }\bibfield  {title} {\enquote {\bibinfo {title} {{Input and
  output in damped quantum systems: Quantum stochastic differential equations
  and the master equation}},}\ }\href {\doibase 10.1103/PhysRevA.31.3761}
  {\bibfield  {journal} {\bibinfo  {journal} {Physical Review A}\ }\textbf
  {\bibinfo {volume} {31}},\ \bibinfo {pages} {3761--3774} (\bibinfo {year}
  {1985})}\BibitemShut {NoStop}%
\bibitem [{\citenamefont {Klimov}\ \emph {et~al.}(2018)\citenamefont {Klimov},
  \citenamefont {Kelly}, \citenamefont {Chen}, \citenamefont {Neeley},
  \citenamefont {Megrant}, \citenamefont {Burkett}, \citenamefont {Barends},
  \citenamefont {Arya}, \citenamefont {Chiaro}, \citenamefont {Chen} \emph
  {et~al.}}]{klimov2018fluctuations}%
  \BibitemOpen
  \bibfield  {author} {\bibinfo {author} {\bibfnamefont {PV}~\bibnamefont
  {Klimov}}, \bibinfo {author} {\bibfnamefont {J}~\bibnamefont {Kelly}},
  \bibinfo {author} {\bibfnamefont {Z}~\bibnamefont {Chen}}, \bibinfo {author}
  {\bibfnamefont {M}~\bibnamefont {Neeley}}, \bibinfo {author} {\bibfnamefont
  {A}~\bibnamefont {Megrant}}, \bibinfo {author} {\bibfnamefont
  {B}~\bibnamefont {Burkett}}, \bibinfo {author} {\bibfnamefont
  {R}~\bibnamefont {Barends}}, \bibinfo {author} {\bibfnamefont
  {K}~\bibnamefont {Arya}}, \bibinfo {author} {\bibfnamefont {B}~\bibnamefont
  {Chiaro}}, \bibinfo {author} {\bibfnamefont {Yu}~\bibnamefont {Chen}},  \emph
  {et~al.},\ }\bibfield  {title} {\enquote {\bibinfo {title} {Fluctuations of
  energy-relaxation times in superconducting qubits},}\ }\href@noop {}
  {\bibfield  {journal} {\bibinfo  {journal} {Physical review letters}\
  }\textbf {\bibinfo {volume} {121}},\ \bibinfo {pages} {090502} (\bibinfo
  {year} {2018})}\BibitemShut {NoStop}%
\bibitem [{\citenamefont {Bergeal}\ \emph {et~al.}(2010)\citenamefont
  {Bergeal}, \citenamefont {Schackert}, \citenamefont {Metcalfe}, \citenamefont
  {Vijay}, \citenamefont {Manucharyan}, \citenamefont {Frunzio}, \citenamefont
  {Prober}, \citenamefont {Schoelkopf}, \citenamefont {Girvin},\ and\
  \citenamefont {Devoret}}]{bergeal2010phase}%
  \BibitemOpen
  \bibfield  {author} {\bibinfo {author} {\bibfnamefont {N}~\bibnamefont
  {Bergeal}}, \bibinfo {author} {\bibfnamefont {F}~\bibnamefont {Schackert}},
  \bibinfo {author} {\bibfnamefont {M}~\bibnamefont {Metcalfe}}, \bibinfo
  {author} {\bibfnamefont {R}~\bibnamefont {Vijay}}, \bibinfo {author}
  {\bibfnamefont {V.~E.}\ \bibnamefont {Manucharyan}}, \bibinfo {author}
  {\bibfnamefont {L}~\bibnamefont {Frunzio}}, \bibinfo {author} {\bibfnamefont
  {D.~E.}\ \bibnamefont {Prober}}, \bibinfo {author} {\bibfnamefont {R.~J.}\
  \bibnamefont {Schoelkopf}}, \bibinfo {author} {\bibfnamefont {S.~M.}\
  \bibnamefont {Girvin}}, \ and\ \bibinfo {author} {\bibfnamefont {M.~H.}\
  \bibnamefont {Devoret}},\ }\bibfield  {title} {\enquote {\bibinfo {title}
  {Phase-preserving amplification near the quantum limit with a josephson ring
  modulator},}\ }\href@noop {} {\bibfield  {journal} {\bibinfo  {journal}
  {Nature}\ }\textbf {\bibinfo {volume} {465}},\ \bibinfo {pages} {64--68}
  (\bibinfo {year} {2010})}\BibitemShut {NoStop}%
\bibitem [{\citenamefont {Opremcak}\ \emph {et~al.}(2018)\citenamefont
  {Opremcak}, \citenamefont {Pechenezhskiy}, \citenamefont {Howington},
  \citenamefont {Christensen}, \citenamefont {Beck}, \citenamefont {Leonard},
  \citenamefont {Suttle}, \citenamefont {Wilen}, \citenamefont {Nesterov},
  \citenamefont {Ribeill} \emph {et~al.}}]{opremcak2018measurement}%
  \BibitemOpen
  \bibfield  {author} {\bibinfo {author} {\bibfnamefont {A}~\bibnamefont
  {Opremcak}}, \bibinfo {author} {\bibfnamefont {I.~V.}\ \bibnamefont
  {Pechenezhskiy}}, \bibinfo {author} {\bibfnamefont {C.}~\bibnamefont
  {Howington}}, \bibinfo {author} {\bibfnamefont {B.~G.}\ \bibnamefont
  {Christensen}}, \bibinfo {author} {\bibfnamefont {M.~A.}\ \bibnamefont
  {Beck}}, \bibinfo {author} {\bibfnamefont {E}~\bibnamefont {Leonard}},
  \bibinfo {author} {\bibfnamefont {J}~\bibnamefont {Suttle}}, \bibinfo
  {author} {\bibfnamefont {C}~\bibnamefont {Wilen}}, \bibinfo {author}
  {\bibfnamefont {K.~N.}\ \bibnamefont {Nesterov}}, \bibinfo {author}
  {\bibfnamefont {G.~J.}\ \bibnamefont {Ribeill}},  \emph {et~al.},\ }\bibfield
   {title} {\enquote {\bibinfo {title} {Measurement of a superconducting qubit
  with a microwave photon counter},}\ }\href@noop {} {\bibfield  {journal}
  {\bibinfo  {journal} {Science}\ }\textbf {\bibinfo {volume} {361}},\ \bibinfo
  {pages} {1239--1242} (\bibinfo {year} {2018})}\BibitemShut {NoStop}%
\bibitem [{\citenamefont {Kono}\ \emph {et~al.}(2018)\citenamefont {Kono},
  \citenamefont {Koshino}, \citenamefont {Tabuchi}, \citenamefont {Noguchi},\
  and\ \citenamefont {Nakamura}}]{kono2018quantum}%
  \BibitemOpen
  \bibfield  {author} {\bibinfo {author} {\bibfnamefont {Shingo}\ \bibnamefont
  {Kono}}, \bibinfo {author} {\bibfnamefont {Kazuki}\ \bibnamefont {Koshino}},
  \bibinfo {author} {\bibfnamefont {Yutaka}\ \bibnamefont {Tabuchi}}, \bibinfo
  {author} {\bibfnamefont {Atsushi}\ \bibnamefont {Noguchi}}, \ and\ \bibinfo
  {author} {\bibfnamefont {Yasunobu}\ \bibnamefont {Nakamura}},\ }\bibfield
  {title} {\enquote {\bibinfo {title} {Quantum non-demolition detection of an
  itinerant microwave photon},}\ }\href@noop {} {\bibfield  {journal} {\bibinfo
   {journal} {Nature Physics}\ }\textbf {\bibinfo {volume} {14}},\ \bibinfo
  {pages} {546--549} (\bibinfo {year} {2018})}\BibitemShut {NoStop}%
\bibitem [{\citenamefont {Besse}\ \emph {et~al.}(2018)\citenamefont {Besse},
  \citenamefont {Gasparinetti}, \citenamefont {Collodo}, \citenamefont
  {Walter}, \citenamefont {Kurpiers}, \citenamefont {Pechal}, \citenamefont
  {Eichler},\ and\ \citenamefont {Wallraff}}]{besse2018single}%
  \BibitemOpen
  \bibfield  {author} {\bibinfo {author} {\bibfnamefont {Jean-Claude}\
  \bibnamefont {Besse}}, \bibinfo {author} {\bibfnamefont {Simone}\
  \bibnamefont {Gasparinetti}}, \bibinfo {author} {\bibfnamefont {Michele~C}\
  \bibnamefont {Collodo}}, \bibinfo {author} {\bibfnamefont {Theo}\
  \bibnamefont {Walter}}, \bibinfo {author} {\bibfnamefont {Philipp}\
  \bibnamefont {Kurpiers}}, \bibinfo {author} {\bibfnamefont {Marek}\
  \bibnamefont {Pechal}}, \bibinfo {author} {\bibfnamefont {Christopher}\
  \bibnamefont {Eichler}}, \ and\ \bibinfo {author} {\bibfnamefont {Andreas}\
  \bibnamefont {Wallraff}},\ }\bibfield  {title} {\enquote {\bibinfo {title}
  {Single-shot quantum nondemolition detection of individual itinerant
  microwave photons},}\ }\href@noop {} {\bibfield  {journal} {\bibinfo
  {journal} {Physical Review X}\ }\textbf {\bibinfo {volume} {8}},\ \bibinfo
  {pages} {021003} (\bibinfo {year} {2018})}\BibitemShut {NoStop}%
\bibitem [{\citenamefont {Johansson}\ \emph {et~al.}(2013)\citenamefont
  {Johansson}, \citenamefont {Nation},\ and\ \citenamefont
  {Nori}}]{johansson2013}%
  \BibitemOpen
  \bibfield  {author} {\bibinfo {author} {\bibfnamefont {J.~R.}\ \bibnamefont
  {Johansson}}, \bibinfo {author} {\bibfnamefont {P.~D.}\ \bibnamefont
  {Nation}}, \ and\ \bibinfo {author} {\bibfnamefont {Franco}\ \bibnamefont
  {Nori}},\ }\bibfield  {title} {\enquote {\bibinfo {title} {{QuTiP 2: A Python
  framework for the dynamics of open quantum systems}},}\ }\href {\doibase
  10.1016/j.cpc.2012.11.019} {\bibfield  {journal} {\bibinfo  {journal}
  {Computer Physics Communications}\ }\textbf {\bibinfo {volume} {184}},\
  \bibinfo {pages} {1234--1240} (\bibinfo {year} {2013})},\ \Eprint
  {http://arxiv.org/abs/1211.6518} {arXiv:1211.6518} \BibitemShut {NoStop}%
\end{thebibliography}%

\end{document}